# The "glue" effect of turbulence in the evolution of an insulated two-layer quasi-geostrophic flow

F. Crisciani [1], R. Mosetti [2], R. Purini [3]

[1] Dip. di Fisica dell'Università di Trieste, Trieste, Italy.
[2] OGS (Istituto Nazionale di Oceanografia e di Geofisica Sperimentale) – Trieste, Italy e Dip. di Matematica e Geoscienze dell'Università di Trieste, Trieste, Italy.
[3] Consiglio Nazionale delle Ricerche – ISAC - Roma, Italy.

**Abstract.** A two-layer quasi-geostrophic flow is quite insulated from the surrounding fluid, while the layers interact each other by means of the modulation of the interface between them and of the turbulence affecting the layers in the proximity of the interface. In this framework, the time evolution of the circulation of the flow along the lateral boundaries (upper and lower) is investigated under the assumption of arbitrary initial conditions. The analytic solution shows the glue-like behaviour of turbulence, which is parameterized in the standard way (Pedosky, 1996): initially the glue is wet and each layer evolves almost independently of the other while, for times far enough from that initial, the glue hardens and the system converges to a single-layer one. Indeed, the circulation of the thinner layer tends to adjust itself in a parallel way with respect to the thicker layer and the final circulation of the consequent single-layer system is given by the average of the initial circulations, each being weighted by the thickness of the related layer.

## 1. Introduction

Turbulence is a field of physics that pertains the fluid-dynamics, the magneto-hydro dynamics of plasma (the so called fourth state of the matter) and the astrophysics. Its complexity is related to the inherent chaotic turbulent motions as the famous paper of Ruelle and Takens pointed out for hydrodynamical systems (Ruelle and Takens, 1971). As a consequence, the investigations on turbulence are mainly based on the numerical integration of the equations describing hydrodynamical systems with sophisticated features (see, for instance, Vallis, 2006, and the references quoted there). To avoid that, in a purely didactic context, computational intricacies mask the fundamental aspect of this phenomenon, i.e. the tendency toward homogenization of the whole fluid body, the author's idea is to focus the homogenizing effect of turbulence by resorting to a (highly idealized) fully insulated two-layer model of a quasi-geostrophic fluid in a bounded and simply connected domain, where turbulence is parameterized according to Pedlosky (1996). Owing to the simplicity of the model which allows a fully analytic treatment, layer's interaction can be described in terms of the two circulations (circulation = line integration of the vector velocity over a given path) along the boundary of the domain, which turn to undergo a strange attractor leading to a final homogeneus circulation of a single fluid layer. With this task in mind, section 2 presents shortly the starting quasi-geostrophic system, section 3 introduces the circulations of both layers thus obtaining a linear dynamical system which solution shows the 'glue' effect of turbulence on the dynamics of the two layers. Par. 4 shows further properties and gives also a quantitative solution for a realistic geo-fluid having the upper layer thinner than the lower one. Some final remarks conclude the paper.

## 2. Governing equations of the model

In this section a special, unforced version of the two-layer quasi-geostrophic model is derived (see Cavallini and Crisciani (2013) for full details). With reference to the $\beta$-plane approximation and in standard notation, in both layers the non dimensional vorticity equation has the form

$$\frac{\partial}{\partial t}\nabla^2\psi + J(\psi,\nabla^2\psi) + \beta\frac{\partial\psi}{\partial x} = \frac{\partial w_1}{\partial z} \tag{2.1}$$

Each layer covers the bounded and simply connected fluid domain $A$ of the $\beta$-plane, whose boundary is denoted with $\partial A$. The whole system is vertically included into the interval $0 \leq z \leq 1$ and the volume $V = A \times (0 \leq z \leq 1)$ is assumed to be a material volume of fluid. The upper layer extends from the rigid lid in $z = 1$, down to the impermeable interface $z = z_i$ where

$$z_i = \bar{z}_i + F(\psi^{II} - \psi^{I}) \tag{2.2}$$

With reference to (2.2), the constant $1 - \bar{z}_i$ is the mean thickness of the upper layer $(0 < \bar{z}_i < 1)$, while $F(\psi^{II} - \psi^{I})$ is the fluctuating part of $z_i$, where $F$ is the rotational Froude number, $\psi^{I}$ is the stream function of the upper layer and $\psi^{II}$ is that of the lower layer. Because of mass conservation and of interface's impermeability

$$\int_A (\psi^{II} - \psi^{I}) dx\, dy = 0 \tag{2.3}$$

The lower layer extends from $z = z_i$ down to the flat bottom in $z = 0$ and has mean thickness $\bar{z}_i$. Hypotheses of a rigid lid and of a flat bottom imply

$$w(z = 1) = w(z = 0) = 0 \tag{2.4}$$

at every level of approximation. Unlike (2.4), the deformation of the interface makes rise to the a-geostrophic vertical velocity $w_1 = w/\varepsilon$ ( $\varepsilon$ is the Rossby number) involving the fluid in both the layers and given by

$$w_1(z = z_i) = F\left[\frac{\partial}{\partial t}(\psi^{II} - \psi^{I}) + J(\psi^{I},\psi^{II})\right] + \frac{\sqrt{E_v}}{2\varepsilon}\nabla^2(\psi^{I} - \psi^{II}) \tag{2.5}$$

where $E_v$ is the vertical Ekman number. The first term at the r.h.s. of (2.5) is the kinematic effect due to the deformation of the interface while the second one represents that of unresolved eddies coupling one layer with the other. One of the simplest realizations of this coupling is a drag law proportional to the velocity difference $\mathbf{u}^{I} - \mathbf{u}^{II}$ between the two layers. Within the quasi-geostrophic approximation $\mathbf{u}^I - \mathbf{u}^{II} = \hat{\mathbf{k}} \times \nabla(\psi^I - \psi^{II})$ and hence the last term at the r.h.s. of (2.5) is explained noting that (Pedlosky, 1996)

$$\hat{\mathbf{k}} \cdot \left\{ \nabla \times (\mathbf{u}^I - \mathbf{u}^{II}) \right\} = \hat{\mathbf{k}} \cdot \left\{ \nabla \times \left[ \hat{\mathbf{k}} \times \nabla(\psi^I - \psi^{II}) \right] \right\} = \nabla^2(\psi^I - \psi^{II}) \qquad (2.6)$$

Now, vertical integration of (2.1) over the thickness of the upper layer, with $\psi^I$ in place of $\psi$, results in the equation

$$\frac{\partial}{\partial t}\nabla^2\psi^I + J(\psi^I, \nabla^2\psi^I) + \beta\frac{\partial \psi^I}{\partial x} = \frac{1}{1-\bar{z}_i}\left[w_I(z=1) - w_I(z=z_i)\right] \qquad (2.7)$$

where $1-\bar{z}_i$ is an appropriate approximation of the actual thickness. By using the first equation of (2.4), (2.5) and setting in short $r = \frac{\sqrt{E_v}}{2\varepsilon}$, equation (2.7) can be restated as

$$\left(\frac{\partial}{\partial t} + \frac{\partial \psi^I}{\partial x}\frac{\partial}{\partial y} - \frac{\partial \psi^I}{\partial y}\frac{\partial}{\partial x}\right)\left[\nabla^2\psi^I - \frac{F}{1-\bar{z}_i}(\psi^I - \psi^{II})\right] + \beta\frac{\partial \psi^I}{\partial x} = -\frac{r}{1-\bar{z}_i}\nabla^2(\psi^I - \psi^{II}) \qquad (2.8)$$

Quite analogously, vertical integration of (2.1) over the thickness of the lower layer, with $\psi^{II}$ in place of $\psi$, results in the equation

$$\left(\frac{\partial}{\partial t} + \frac{\partial \psi^{II}}{\partial x}\frac{\partial}{\partial y} - \frac{\partial \psi^{II}}{\partial y}\frac{\partial}{\partial x}\right)\left[\nabla^2\psi^{II} - \frac{F}{\bar{z}_i}(\psi^{II} - \psi^I)\right] + \beta\frac{\partial \psi^{II}}{\partial x} = -\frac{r}{\bar{z}_i}\nabla^2(\psi^{II} - \psi^I) \qquad (2.9)$$

Equations (2.8) and (2.9) govern the evolution of the fluid layers of the material volume $V$ under the assumption that they are dynamically insulated with respect to the surrounding ambient. At the same times, the layers interact each other through the vertical velocity (2.5) of the interface.

### 3. The geostrophic circulation in the layers

To explain the glue effect of turbulence, the circulation of the geostrophic currents in each layer along the boundary of the fluid domain is introduced by means of positions

$$C^I(t) = \oint_{\partial A} \mathbf{u}^I(s,t) \cdot \hat{\mathbf{t}}\, ds \qquad (3.1)$$

and

$$C^{II}(t) = \oint_{\partial A} \mathbf{u}^{II}(s,t) \cdot \hat{\mathbf{t}}\, ds \qquad (3.2)$$

Because $\mathbf{u}^I = \hat{\mathbf{k}} \times \nabla \psi^I$ and $\mathbf{u}^{II} = \hat{\mathbf{k}} \times \nabla \psi^{II}$, equations (3.1) and (3.2) are equivalent to

$$C^I(t) = \int_A \nabla^2 \psi^I \, dx\, dy \tag{3.3}$$

and

$$C^{II}(t) = \int_A \nabla^2 \psi^{II} \, dx\, dy \tag{3.4}$$

respectively. The evolution equations of (3.1) and (3.2) can be inferred by using (3.3) and (3.4) after integration of (2.8) and (2.9) over the fluid domain $A$ with the aid of the Reynolds' transport theorem (see, for instance, Salby (1996)). Recalling also (2.3) and noting that $\int_A \frac{\partial \psi^I}{\partial x} dxdy = \int_A \frac{\partial \psi^{II}}{\partial x} dxdy = 0$, integration yields the linear dynamic system

$$\begin{cases} \dfrac{dC^I}{dt} = -\dfrac{r}{1-\overline{z}_i}\left(C^I - C^{II}\right) \\ \dfrac{dC^{II}}{dt} = \dfrac{r}{\overline{z}_i}\left(C^I - C^{II}\right) \end{cases} \tag{3.5}$$

To single out a unique solution, system (3.5) must be supplemented with initial conditions

$$C^I(0) = C_0^I, \quad C^{II}(0) = C_0^{II} \tag{3.6}$$

where $C_0^I$ and $C_0^{II}$ are prescribed constants. The solution of problem (3.5), (3.6) is

$$\begin{aligned} C^I(t) &= C_0^I + \left(C_0^I - C_0^{II}\right)\overline{z}_i \left\{\exp\left[-\frac{r}{\overline{z}_i(1-\overline{z}_i)}t\right] - 1\right\} \\ C^{II}(t) &= C_0^{II} + \left(C_0^I - C_0^{II}\right)(\overline{z}_i - 1)\left\{\exp\left[-\frac{r}{\overline{z}_i(1-\overline{z}_i)}t\right] - 1\right\} \end{aligned} \tag{3.7}$$

In particular, solution (3.7) implies

$$\lim_{t\to+\infty} C^I(t) = \lim_{t\to+\infty} C^{II}(t) = \left(1-\overline{z}_i\right)C_0^I + \overline{z}_i\, C_0^{II} \tag{3.8}$$

According to (38), *both circulations (3.1) and (3.2) converge asymptotically to the same limit*, given by the average of $C_0^I$ and $C_0^{II}$ with weights $1-\overline{z}_i$ and $\overline{z}_i$, respectively. Thus, turbulence looks like the effect of a glue between the layers; in fact, as long as the glue is wet (very short time after that initial), circulations are independent each other owing to the arbitrariness of the initial conditions (3.6) while, after glue hardening, circulations become strictly concordant (as (3.8) shows) and converge to a single-layer system.

Relationship (3.8), in which the limit is different of zero unless very special initial conditions are considered, relies on the singularity of the matrix

$$M = \begin{pmatrix} -r/(1-\overline{z}_i) & r/(1-\overline{z}_i) \\ r/\overline{z}_i & -r/\overline{z}_i \end{pmatrix}$$

associated to (3.5). In fact $det(M) = 0$ and, as a consequence, solution (3.7) is given by the superposition of an exponentially decreasing function plus a constant function. The singularity can be eliminated if the request of an insulated system is released and bottom friction is introduced in (2.9) by adding a term of the kind $-\dfrac{s}{\overline{z}_i}\nabla^2 \psi^{II}$ $(s > 0)$ at its r.h.s. In this case, matrix above changes into

$$\tilde{M} = \begin{pmatrix} -r/(1-\overline{z}_i) & r/(1-\overline{z}_i) \\ r/\overline{z}_i & -(r+s)/\overline{z}_i \end{pmatrix}$$

so $det(\tilde{M}) = \dfrac{rs}{\overline{z}_i(1-\overline{z}_i)} > 0$ and, hence, after little algebra one concludes that $\lim\limits_{t \to +\infty} C^I(t) = \lim\limits_{t \to +\infty} C^{II}(t) = 0$ unlike (3.8).

### 4. Further properties of (3.1) and (3.2)

System (3.5) implies

$$\frac{dC^I}{dt} \bigg/ \frac{dC^{II}}{dt} < 0 \qquad (4.1)$$

and

$$\frac{|dC^I/dt|}{|dC^{II}/dt|} = \frac{\overline{z}_i}{1-\overline{z}_i} \qquad (4.2)$$

Inequality (4.1) shows that *circulations approach the asymptotic limit with opposite growth rates*. In fact, the asymptotic limit at the r.h.s. of (3.8) is necessarily included between initial conditions (3.6) so one of (3.1), (3.2) increases in time while the other decreases.

Equation (4.2) shows that *the intensity of the growth rate of the circulation is inversely proportional to the thickness of the layer*. In other words, the thicker layer approaches more slowly than that thinner to the asymptotic single-layer system.

The monotonic character of solution (3.7) implies that $C^I(t)$ *reverts once its sign in the course of time* if and only if $C^I(0)$ and $\lim\limits_{t \to +\infty} C^I(t)$ have discordant signs, that is to say *if and only if*

$$C_0^I \left[ (1-\overline{z}_i)C_0^I + \overline{z}_i C_0^{II} \right] < 0 \qquad (4.3)$$

Analogously, $C^{II}(t)$ *reverts once its sign if and only if*

$$C_0^{II}\left[(1-\bar{z}_i)C_0^I + \bar{z}_i\ C_0^{II}\right] < 0 \tag{4.4}$$

However, *the inversion of the circulation in both the layers is not possible, even in deferred times*. In fact each of inequalities (4.3) and (4.4) imply

$$C_0^I C_0^{II} < 0 \tag{4.5}$$

so $C_0^I$ and $C_0^{II}$ should have concordant signs because of (4.3) and (4.4) but also discordant signs because of (4.5).

Example. Consider a hypothetical two-layer system with the upper layer thinner than that lower, say

$$1-\bar{z}_i = 0.3,\ \bar{z}_i = 0.7 \tag{4.6}$$

and assume $r = 1$. Let initial circulations be opposite and such that

$$C_0^I = -1,\ C_0^{II} = 1 \tag{4.7}$$

Owing to (4.6) and (4.7), equation (3.8) states both circulations converge asymptotically to the value $0.4$. The upper (thinner) layer approaches this value faster than that lower; in fact, substitution of (4.6) into (4.2) yields

$$\left|\frac{dC^I}{dt}\right| \cong 2.3 \left|\frac{dC^{II}}{dt}\right| \tag{4.8}$$

Moreover, the l.h.s. of inequality (4.3) is equal to $-0.4$, thus showing that, unlike the lower layer, the circulation in the upper layer changes it sign in the course of time. All this results in the plots of (3.7) that, here, takes the form

$$\begin{aligned} C^I(t) &\cong -1 - 1.4\left[\exp(-4.8\ t) - 1\right] \\ C^{II}(t) &\cong 1 + 0.6\left[\exp(-4.8\ t) - 1\right] \end{aligned} \tag{4.9}$$

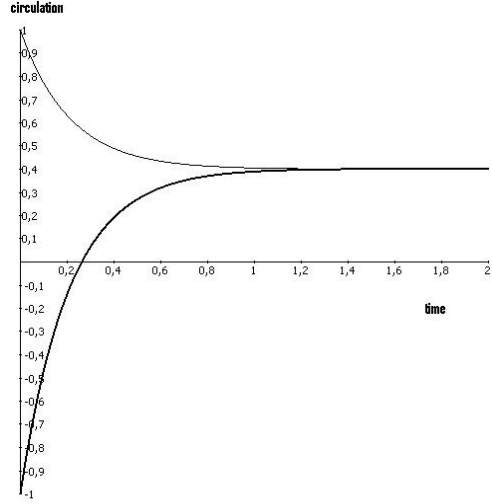

Time evolution of $C^I(t)$ (lower plot) and of $C^{II}(t)$ (upper plot), according to (4.9) from the initial time $t = 0$ up to the almost exact coincidence of the quantities above. Note the inversion of the sign of $C^I(t)$ in $t \cong 0.26$.

**Remark.**

The dynamical system (3.5) is left unaltered even if the lower layer is bounded from below by a topographic modulation and its vertical extension is

$$\eta(x,y) \leq z \leq z_i(x,y,t) \qquad (4.10)$$

in place of $0 \leq z \leq z_i(x,y,t)$. In fact, owing to (4.10), equations (2.4) are modified as

$$w(z=1)=0, \; w(z=\eta)=\varepsilon J(\psi^{II},\eta) \qquad (4.11)$$

and, because of (4.11), equation (2.9) is supplemented by a further term as follows

$$\left(\frac{\partial}{\partial t}+\frac{\partial \psi^{II}}{\partial x}\frac{\partial}{\partial y}-\frac{\partial \psi^{II}}{\partial y}\frac{\partial}{\partial x}\right)\left[\nabla^2 \psi^{II}-\frac{F}{z_i}(\psi^{II}-\psi^I)+\eta\right]+\beta\frac{\partial \psi^{II}}{\partial x}=-\frac{r}{z_i}\nabla^2(\psi^{II}-\psi^I) \qquad (4.12)$$

However, the identity

$$\int_A J\left(\psi^{II}, \eta\right) dxdy = 0 \qquad (4.13)$$

holds true whatever the differentiable function $\eta(x,y)$ may be (Cavallini and Crisciani, 2013), and therefore integration of (4.12) over the fluid domain $A$ with the aid of (4.13) yields the same result as (2.9), so (3.5) in any case is obtained.